\documentstyle[12pt]{article} 
\newcommand{\be}{\begin{equation}}
\newcommand{\ee}{\end{equation}}
\newcommand{\beq}{\begin{eqnarray}}
\newcommand{\eeq}{\end{eqnarray}}
\newcommand{\bear}{\begin{array}}
\newcommand{\ear}{\end{array}}

\begin{document}
\title{Time In Quantum Gravity}
\author{S.Biswas, A.Shaw and B.Modak}
\date{}
\maketitle
\begin{center}
 Department of Physics\\
 University of Kalyani\\
 P.O.- Kalyani, Dst.- Nadia\\
 West Bengal (India)\\
 Pin. - 741235
\end{center}
\begin{abstract}
The Wheeler-DeWitt equation in quantum gravity is timeless in character. 
In order to discuss quantum to classical transition of the universe, one uses 
a time prescription in quantum gravity to obtain a time contained description
starting from Wheeler-DeWitt equation and WKB ansatz for the WD wavefunction.
The approach has some drawbacks. In this work, we obtain the time-contained
Schroedinger-Wheeler-DeWitt equation without using the WD equation and the WKB
ansatz for the wavefunction. We further show that a Gaussian ansatz for SWD 
wavefunction is consistent with the Hartle-Hawking or wormhole dominance 
proposal boundary condition. We thus find an answer to the small scale boundary
conditions.
 
\end{abstract}
\newpage
\section{\bf{Introduction}}
A basic feature of quantum cosmology is that the universe starts with a quantum
character being dominated by quantum uncertainty and eventually it then becomes
large and completely classical. In quantum cosmology the universe is described by
a wavefunction $\Psi $ which satisfies the equation
\be
\hat{H}\Psi=0
\ee
where $\hat{H}$ is the Hamiltonian operator. The equation (1) is known as the
Wheeler-DeWitt (WD) equation. Equation (1) when compared with the Schroedinger
equation in quantum mechanics
\be
i\hbar{\partial \over \partial t}= \hat{H}\Psi ,
\ee
reveals that there is no "time" in quantum gravity and this is commonly referred
to as `the problem of time' in quantum gravity. It is now accepted as a broad
consensus [1,2,3,4]  that the time in quantum gravity has an intrinsic character.
Recent trends suggest that one achieves an equation like (2) through a prescription
of time. The problem with equation (1) is not to find solutions but to find
a proper boundary condition that will not disturb the basic aspect of inflationary
cosmology. At present we have three boundary condition proposals. These are : (i).
Hartle-Hawking no boundary proposal [5] (ii). quantum tunneling proposal [6] and less
commonly known (iii). wormhole dominance proposal [7]. The third boundary condition
is more general in the sense that the proposals (i) and (ii) follow from (iii)
when the respective boundary conditions are introduced in it. The first two proposals produce
wavefunctions that are not normalized and have to rely on the concept of
conditional probability [8] for an interpretation of the wavefunction. The 
proposal (iii) obtains wavefunction which is normalisable and the probabilistic
interpretation remains quite sensible, and workable as in ordinary quantum mechanics.
The problem with (2) is also to choose suitable initial conditions and to obtain 
a reasonable connection with the three boundary conditions. In most works an 
equation like (2) is derived from (1) and the equation is called 
Schroedinger-Wheeler-DeWitt (SWD) equation [1,2,3,4].
\par 
As mentioned the major problem in quantum gravity is not to find solutions of the WD and SWD
equations but to obtain suitable initial conditions such that the inflationary 
scenario for the early universe and fruits emerging out of it are not changed. It
is now accepted that the inflation provides a natural mechanism for structure 
formation and its origin is traced back to the quantum fluctuations in early 
universe. These quantum fluctuations are related to a scalar field $\phi$ in 
phase transition model and to the geometry itself in Starobinsky's spontaneous 
transition model [9]. The idea of quantum universe necessitates an interpretation of the 
wavefunction. For the orthodox ``Copenhagen interpretation'' one requires an 
external classical observer but for a universe there is no observer external to 
it. The success of classical Einstein equation along with the classical spacetime is 
a reality, so we need along with an interpretation of the wavefunction, also a 
mechanism from microscopic to macroscopic reduction. More specifically, we need 
a mechanism from quantum to classical transitions. There have been many 
discussions for a unified dynamics for microscopic and macroscopic systems [10]. 
Now it is known that classical properties emerges through the interactions of 
the variables describing a quantum system such that the configuration variables are 
divided in some way into macroscopic variables $M$ and microscopic variables $Q$ 
and quantum interference effects are suppressed by averaging out the microscopic 
variations not distinguished by the associated observable. This process is known 
as decoherence [11,12].
\par   
In the context of quantum gravity, the Hamiltonian constraint leads to the 
timeless WD equation and recovery of semiclassical time is carried out using two 
main approaches. In one approach [13,14] a variable $t$ (depending on the 
original position and momenta) whose conjugate momenta occurs linearly in the 
Hamiltonian $H$ is brought to a form 
\be
H=H_r+p_t\,,
\ee
through a canonical transformation. The quantization 
$p_t\rightarrow -i\hbar {\partial \over {\partial t}}$
then brings (3) to the form (2) and obtains SWD equation from the Hamiltonian 
constraint $H=0$. This approach succeeds in some cases like cylindrical 
gravitational waves or eternal black holes but its general viability is far from 
clear though the standard Hilbert space of quantum theory can be employed in such
an approach. 
\par
The other approach starts from the WD equation (1) and treats all variables in 
the same footing and tries to identify a sensible concept of time after 
quantization. In this approach (i) the choice for an appropriate Hilbert space 
structure is obscure, (ii) normalization of the wavefunction and probabilistic 
interpretation remain awkward in absence of time and (iii) whether the 
prescription of time parameter is an artifact and is related to Minskowskian time 
are not clear. Though the concept of `conditional probabilities' is enforced for 
an interpretation of the wavefunction, the driving quantum force guaranting the 
validity of superposition principle in early universe and subsequently 
enforcing decoherence remain unclear in the picture.
\par   
The motivation of the present work is to investigate the role of time in quantum
gravity especially to understand the initial conditions of both the WD and SWD 
equations and to obtain an inter-relation between them. In our approach we do 
not enforce any canonical transformation to obtain an equation like (3) and do 
not consider the Wheeler-DeWitt equation to obtain the SWD equation. If we look 
at classical Einstein equation 
\be
G_{\mu\nu}\equiv R_{\mu\nu}- {1\over 2}g_{\mu\nu}R=kT_{\mu\nu}\,,
\ee
we observe that `geometry and matter' get coupled through (4). It is also a 
well known fact that the matter field is quantized and for that reason in equation 
(4) one writes $<T_{\mu\nu}>$ on the right hand side of (4) and treats 
$g_{\mu\nu}$ as classical background. Keeping this spirit of (4) in mind we introduce
Minskowski time $t$ through Schroedinger equation
\be
i\hbar{{\partial \psi}\over {\partial t}}={\hat{H}}_m\psi\,,
\ee
where ${\hat{H}}_m$ is now the matter field Hamiltonian. This $t$ now serves as an 
external label. Without having the gravitational field quantized, we formulate a 
time parameter $\sigma (x)$ such that (5) becomes equivalent to Einstein 
equation with $\sigma(x)=t=const.$ acting as a global parameter. The geometry 
itself acts as a generator of time and manifest only through matter field. We 
discuss this recovery of semiclassical time in section II. In section III we 
discuss the initial conditions for solution of SWD equation and its connection 
to the three boundary condition proposals for the timeless Wheeler-DeWitt 
equation mentioned earlier in the introduction. 
Section IV contains a discussion of `quantum force' originated in the geometry 
through wormhole picture.
\smallskip
\section{\bf{Semiclassical Time in Quantum Gravity}}
We consider a gravitational action with a minimally coupled scalar field $\phi$
in a Friedman-Robertson-Walker (FRW) background
\be
I=M\int{dt\left[-{1\over 2}a{\dot{a}}^2+{{ka}\over 2}+{1\over M}
\{{1\over 2}{\dot{\phi}}^2-V(\phi)\}\;a^3\,\right]}\,,
\ee
where $M={{3\pi}\over {2G}}={{3\pi m_{p}^{2}}\over 2},\, m_p$ being the Planck mass
and $k=0,\pm 1$ for flat, closed and open models respectively. The 
$\left(\bear{c}0\\0\ear\right)$ component of Einstein equation is 
\be
-{M\over 2}({{{\dot{a}}^2}\over a^2}+{k\over a^2})+{1\over 2}
{\dot{\phi}}^2+V(\phi)=0\,.
\ee
Identifying
\be
P_a=-Ma\dot{a}\,,\, P_{\phi}=a^3\dot{\phi}\,,
\ee
(7) gives the Hamiltonian constraint
\be
-{1\over {2M}}({P_{a}^{2}\over a})+{P_{\phi}^{2}\over {2a^3}}
-{M\over 2}ka+a^3V(\phi)=0\,.
\ee
The dynamical equations are
\be
-{{\ddot{a}}\over a}={{\dot{a}}^2\over {2a^2}}+{k\over {2a^2}}+ {3\over M}
\{{1\over 2}{\dot{\phi}}^2-V(\phi)\}\,,
\ee
\be
-\ddot{\phi}=3{\dot{a}\over a}\dot{\phi}+{{\partial V}\over {\partial \phi}}\,.
\ee
The matter field Hamiltonian $H_m$ for the scalar field is
\be
H_m={1\over {2a^3}}P_{\phi}^{2}+a^3V(\phi)
\ee
as if $a^3({1\over 2}{\dot{\phi}}^2+V(\phi))=E$ is the energy of the scalar 
field. We now define an action $S(a,\phi)$ such that
\be
H_m=-{{\partial S}\over {\partial t}}
\ee
and (12) reduces to 
\be
-{{\partial S}\over {\partial t}}={1\over {2a^3}}P_{\phi}^{2}+a^3V(\phi)\,.
\ee
This $t$ is obviously a Newtonian time and acts as an external label. 
Seemingly it appears that (14) has no connection with the gravitational field.
Using the Hamiltonian constraint (9), we write (14) as
\be
-{{\partial S}\over {\partial t}}=+{1\over {2M}}{P_{a}^{2}\over a}
+{k\over 2}Ma\,.
\ee
If we quantize in standard way with 
$P_i=-i\hbar {\partial\over {\partial q_i}}$,
(14) and (15) are added we get the Wheeler-DeWitt equation and the time disappears from
the equation and this is the problem of time in quantum gravity. Our view is that
quantization is permitted in (14) but not in (15) as if (15) represent the
classical Einstein equation whereas (14) with 
$p_t={{\partial S}\over {\partial t}}=-i\hbar {\partial\over {\partial t}}$
and $P_\phi=-i\hbar {\partial\over {\partial \phi}}$ acting as quantum equation.
It seems as if (14) and (15) have no dynamical content.
\par
We define therefore a time evolution parameter $\sigma$
\be
{\partial\over {\partial \sigma}}=
{{\partial H}\over {\partial P_a}} {\partial\over {\partial a}}
+{{\partial H}\over {\partial P_\phi}} {\partial\over {\partial \phi}}
-{{\partial H}\over {\partial a}} {\partial\over {\partial P_a}}
-{{\partial H}\over {\partial \phi}} {\partial\over {\partial P_\phi}}\,.
\ee
Using (9) and (16) one finds
\beq
{\partial\over {\partial \sigma}}&=&-{1\over {Ma}}
({{\partial S}\over {\partial a}}){\partial\over {\partial a}}
+{1\over a^3}({{\partial S}\over {\partial \phi}}){\partial\over {\partial \phi}}\nonumber \\
&+&\left[{{kM}\over 2} -
{{({{\partial S}\over {\partial a}})^2}\over {2Ma^2}}
+3{{({{\partial S}\over {\partial \phi}})^2}\over {2a^4}}-3a^2V(\phi)\right]
{\partial\over {\partial P_a}}\nonumber \\
&-&a^3({{\partial V}\over {\partial \phi}}){\partial \over {\partial P_\phi}}\,.
\eeq
In view of (14) and (15), we demand that $\sigma$ depends only upon geometry
(i.e., on $a$). This necessitates the co-efficients of 
${\partial\over {\partial P_a}}$ and ${\partial\over {\partial P_\phi}}$ to
become zero in (17). This gives $V(\phi)=0$, and $S$ is a function of $a$ 
only with 
\be
{{({{\partial S_o}\over {\partial a}})^2}\over {2Ma^2}}-{kM\over 2}=0\,.
\ee
The second term vanishes identically. We henceforth denote $S_o=S(a)$. Thus we have
\be
{\partial\over {\partial \sigma}}=-{1\over {Ma}}
{{\partial S_o}\over {\partial a}}{\partial\over {\partial a}}\,.
\ee
Let us suppose that $S$ defined in (14), (15) and (17) is related to $S_o(a)$
by the relation
\be
S(a,\phi)=S_o(a)+S_1(\phi)
\ee
with $S_1(\phi)<<S_o(a)$. 
The reason for such an assumption will be clear as we 
proceed through the text. From (19) we write
\be
{{\partial S}\over {\partial \sigma}}=-{1\over {Ma}}
{{\partial S_o}\over {\partial a}}{{\partial S}\over {\partial a}}\,,
\ee
and using (20) and (21), we find 
\be
{{\partial S}\over {\partial \sigma}}=-
{1\over {Ma}}
({{\partial S}\over {\partial a}})^2
+{1\over {Ma}}
{{\partial S_1}\over {\partial a}}
{{\partial S}\over {\partial a}}\,.
\ee
Because of $S_1(\phi)<<S_o(a)$, we write the second term in (22), using (18) as
\beq
{1\over {Ma}}
{{\partial S_1}\over {\partial a}}
{{\partial S_o}\over {\partial a}}&=&
{1\over {Ma}}
({{\partial S_o}\over {\partial a}})
({{\partial S}\over {\partial a}})-
{1\over {Ma}}
({{\partial S_o}\over {\partial a}})^2\,,\nonumber \\
&=&-{{\partial S}\over {\partial \sigma}}-
{1\over {Ma}}kM^2a^2\,,
\eeq 
so that (22) reduces to 
\be
{{\partial S}\over {\partial \sigma}}=-
{1\over {2Ma}}P_{a}^{2}-{{kMa}\over 2}\,.
\ee
In arriving at (24) we have neglected 
${1\over {Ma}}({{\partial S_1}\over {\partial a}})^2$
term in (23) and is quite obvious in large $a$ region. Comparing (24) with (15) we
find that
\be
{{\partial S}\over {\partial \sigma}}=
{{\partial S}\over {\partial t}}\,.
\ee
The prescription (19) and the condition (25) implants the geometry in the 
quantum structure provided $\sigma=t$ is identified as time. Thus we have 
avoided the quantization of gravitational field. This does not imply that the 
gravitational field is not quantized. It manifests its quantum character only 
through the matter field which is quantized. If we quantize the gravitational 
field through the replacement 
$P_a=-i\hbar{\partial\over {\partial a}}$, 
time will disappear and this is why we get timeless character of the 
Wheeler-DeWitt equation. Now upon quantization with                                                                                             
$P_i=i\hbar{\partial\over {\partial q_i}}$ with $q_i=t,\phi$, we get from (14) 
\be
i\hbar{{\partial \psi}\over {\partial t}}=\left[-{{\hbar^2}\over {2a^3}}
\partial_{\phi}^{2}+a^3V(\phi)\right]\,\psi \,.
\ee
This is the desired Schroedinger-Wheeler-DeWitt equation of quantum gravity and 
works with the standard Hilbert space structure. The derivative ${\partial\over 
{\partial t}}$ is a directional derivative along each of the classical 
spacetime and is viewed as classical `trajectories' in gravitational configuration space. 
In conformity with classical Einstein equation, (26) describes quantized matter 
field in a classical background of spacetime. In other works one starts from the 
WD equation with the ansatz 
\be
\Psi({\it{G}},\phi)\simeq C[{\it{G}}]\,e^{{iS_o[{\it{G}}]}\over \hbar}
\psi [{\it{G}},\phi]\,,
\ee
where ${\it{G}}$ denotes the gravitational field on a three dimensional space,
$\phi$ stands for non-gravitational field and $C$ is a slowly varying prefactor. 
In obtaining (26) from the Wheeler-DeWitt equation, a WKB form has to be assumed 
for $\Psi$ in (1) and
expand $S(a,\phi)$ as
\be
S=MS_o+M^{-1}S_2+\ldots\,.
\ee
Substituting all these in (1) and equating co-efficient of different orders of
$M$ to zero, one finds for $M^2$ order ${{\partial S_o}\over {\partial \phi}}=0,
\,M^1$ order gives source free Hamilton-Jacobi equation
\[ ({a^2\over 2})({{\partial S_o}\over {\partial a}})^2+{{ka^4}\over 2}=0\]
and the prefactor $C({\it{G}})$ is determined through a condition such that
\be
f(a,\phi)=C(a)\exp{({{iS_1}\over \hbar})}
\ee
satisfies (26). As the fundamental equation (1) is linear, it allows arbitrary
superposition of states like (27) and would thus forbid the derivation of (26).
Our approach remains free from all these defects. Our derivation itself suggests, (26)
is valid in semiclassical region because taking
\be
\psi=\exp{\left[{i\over \hbar}S(a,\phi)\right]}\,,
\ee
in (26), we find
\be
-{{\partial S}\over {\partial t}}={1\over {2a^3}}P_{\phi}^{2}+a^3V(\phi)
-{{i\hbar}\over {2a^3}}{{\partial^2 S}\over {\partial \phi^2}}\,,
\ee
which in the limit $\hbar\rightarrow 0$ gives back the classical equation (15).
Thus we conclude that the WD equation remains valid in high curvature region 
(small $a$) and the SWD equation is effective in small curvature region 
(large $a$). The emergence of (31) also implies that somehow the superposition has 
been wiped out during the evolution and this mechanism is known as decoherence. 
It is argued that the seed of this decoherence i.e., the non-occurrence of 
interference terms lies in the early universe. More precisely, initial conditions 
at early time (i.e., near the onset of inflation) somehow regulates 
the behavioural pattern of the wavefunction necessitating decoherence. We now 
discuss this initial condition for the SWD equation (26). The present trend
of investigation concentrates on the quantum to classical transition of the 
universe especially in the light of decoherence mechanism. We discuss it in 
the next section.
\smallskip
\section{\bf{Initial Conditions}}
The Wheeler-DeWitt equation is a quantum equation and the SWD equation is a 
semiclassical equation. It is therefore necessary to prescribe an initial  
condition for (26) so that decoherence can be effective in the framework 
and investigate the nature of initial condition for WD equation that results 
from such a choice, provided the inflation is guaranted in the description.
The usual assumption [4] is that at an early time, the modes are in their 
adiabatic ground state and the initial adiabatic ground state is a Gaussian 
state for the wavefunction since the Gaussian ansatz preserves the Gaussian 
form during time evolution. We will now show that this initial condition is not 
sufficient and requires a condition for the normalization of the wavefunctions. 
For this purpose we start with equation (26) with a form of the potential   
\be
V(\phi)={\lambda\over 2}(1+m^2\phi^2)\,,
\ee
$\lambda$ and $m^2$ being constants, and $m^2$ can also be chosen as negative. 
We take for $\psi$, the Gaussian ansatz 
\be
\psi=N(t)\exp{\left[-{1\over 2}\Omega (t)\right]}\phi^2
\ee
for the ground state of the wavefunction. Inserting (33) in (26) one finds the                                                                     
set of equations
\be
i{d\over {dt}}\ln N={\Omega\over a^3}+\lambda a^3\,,
\ee
\be
-i\dot{\Omega}={{\omega^2-\Omega^2}\over a^3}\,,
\ee
with
\be
\omega^2=m^2\lambda \,.
\ee
It is worthwhile to point out that in (34) and (35)
\be
\dot{\Omega}={{\partial \Omega}\over {\partial t}}\,,
\ee
\be
{1\over 2}{d\over {dt}}\ln N={\partial\over {\partial t}}\ln N
\ee
because of (25) since we would evaluate $N$ considering multiple reflections at 
the turning points arising out of the WD equation such that $N=N(t,\sigma)$. 
Identification of `many fingered' time $\sigma$ as a global parameter leads to the 
choice (38). We now introduce conformal time $\eta$ through the relation 
$dt=a\,d\eta$ to reduce (37) to the form
\be
y''+2{a'\over a}y'+m^2\lambda a^2y=0\,.
\ee
In obtaining (39) we have taken
\be
\Omega=-ia^2{y'\over y}\,,
\ee
in which $y'={{\partial y}\over {\partial \eta}}$. As we require an exponential 
expansion, we solve (39) with $a(\eta)=-{1\over {\sqrt{\lambda}\,\eta}}$. In this 
case (39) reads
\be
y(\eta)=\eta^{{3\over 2}\pm\sqrt{{9\over 4}-m^2}}\,.
\ee
In the limit $m^2<{9\over 4}$ (which is usually assumed to be satisfied in 
inflationary model), where the exponent in (41) can be approximated (taking 
negative sign) as ${1\over {3m^2}}$ so that
\be
\Omega\approx {{im^2a^3\sqrt{\lambda}}\over 3}\,.
\ee
As $\Omega$ is imaginary, the state (33) will not be normalizable. Usually, 
higher order modes of the scalar field are considered to obtain a real part in 
$\Omega$, but here we take a  different procedure. To be consistent with 
standard notation we take $m^2\lambda=m_{o}^{2}$, the mass of the scalar field, 
and $\lambda=H^2$ so that (42) reduces to
\be
\Omega={{im_{o}^{2}a^3}\over {3H}}\,,
\ee  
where $H$ is now the Hubble constant. The imaginary $\Omega$ that contains now 
the mass describes back reaction. Substituting (43) in (34) and taking 
\be
{d\over {dt}}=\sqrt{\lambda}a{d\over {da}}\equiv Ha{d\over {da}}\,,
\ee
we find from (34) 
\be
N=N_o a^{m_{o}^{2}\over {3H}}\exp \left[{-{ia^3H}\over 3}\right]\,,
\ee
so that
\be
\psi=N_o a^{m_{o}^{2}\over {3H}}\exp \left[{-{ia^3H}\over 3}
(1+{1\over 2}m^2\phi^2)\right]\,.
\ee
Since $m^2\phi^2<<1$, we write (46) absorbing a factor 2 in $V(\phi)$ to make
comparison with the standard result [6,7] as
\be
\psi\simeq N_o a^{m_{o}^{2}\over {3H}}\exp \left[{-i\over {3V}}
(a^2V-1)^{3\over 2}\right]\,,
\ee
where we have taken $a^2V>>1$ as expected and 
$\sqrt{V(\phi)}=H(1+{1\over 2}m^2\phi^2)$. From the WD equation, we know that 
$a^2V>>1$ and $a^2V<<1$ regions are respectively termed as classically allowed 
and classically forbidden region. The points $a=0$ and $a={1\over \sqrt{V}}$ are 
the turning points. According to wormhole dominance proposal [7], the normalization 
constant $N_o$ is given by multiple reflections such that
\be
N_o={{\exp{\left[S(a_o,0)\right]}}\over {1-\exp{\left[2S(a_o,0)\right]}}}\,,
\ee
where
\be
S(a_2,a_1)=\left[{{-i(a^2V-1)^{3\over 2}}\over {3V}}\right]_{a_1}^{a_2}\,.
\ee
Evaluating (48) we find 
\be
N_o={{\exp{\left[{1\over {3V}}\right]}}\over {(1-\exp{\left[{2\over {3V}}\right]})}}\,.
\ee
The wavefunction (47) now reads 
\be
\psi=C_1a^{m_{o}^{2}\over {3H}}
\exp\left[{1\over {3V}}(1-i(a^2V-1)^{3\over 2})\right]\,,
\ee
where $C_1$ now refers to $(denominator)^{-1}$ in (50). Continuing in classical 
forbidden region, we get from (51)
\be
\psi(a^2V<1)=C_1a^{m_{o}^{2}\over {3H}}
\exp\left[{1\over {3V}}(1-(1-a^2V)^{3\over 2})\right]\,.
\ee
We see from (52) that as $a\rightarrow 0$, the wavefunction behaves as
\be
\psi\sim\exp({a^2\over 2})\,.
\ee
Thus we find that Eqns.(51)-(53) all represent Hartle-Hawking wavefunction. Thus 
we find that Gaussian ansatz at $a^2V>>1$ correctly reproduce the wavefunction 
corresponding to the wormhole-dominance proposal, at least the Hartle-Hawking 
proposal. Not only the Gaussian ansatz correctly describes the inflation, its 
seed lie in the wormhole-dominance proposal. 
\smallskip
\section{\bf{Discussion}}
With the inception of quantum cosmology and description through the 
Wheeler-DeWitt equation attempts have been made to interpret the wavefunction of 
the universe in terms of classical dynamics (i.e., Einstein equation) and 
probabilistic concept. We have been able to show that a Gaussian ansatz for SWD 
wavefunction correctly simulates the Hartle-Hawking wavefunction and is 
normalized according to our prescription. The prescription obtains the 
normalization constant through the contribution of wormholes  
to the wavefunction and it contributes mainly around the zero scale factor 
region. In otherwords, repeated reflections between the turning points and 
superposition of states like $\exp{(iS)}$ and $\exp{(-iS)}$ are basically the 
quantum character and owe its origin to the wormholes. Our work shows that the 
times $`t'$ and $\sigma$ become equal at the semiclassical region and begin to 
differ as we approach more and more to the classically forbidden region. We observe 
that the quantization becomes worthy only through the matter Lagrangian i.e., 
the Newtonian time emerges through the matter field Hamiltonian in SWD equation. 
This is in conformity with classical Einstein equation where in the righthand 
side we use $<T_{\mu\nu}>$ to obtain the classical description i.e., the 
geometry acts as a ``guidance field'' for matter. Akin to Madelung [15] and 
Bohm [16] in our approach the initial positions are random and the quantum force 
generating the randomness arises from wormhole contribution. In view of some 
apathy towards the wormhole philosophy, we like to mention that the quantum 
force generated at the initial stage (i.e., repeated reflections at the turning 
points and the superposition) finds an interpretation through wormhole 
contributions. This also guarantees the universal validity of superposition 
principle. In otherwords, the geometry in the early universe initiates the 
quantum randomness and with the beginning of nucleation and inflation, the 
Newtonian time emerges through the SWD equation such that $S_1<<S_o$ and 
$S_o=S_o(a)$, an aspect relevant to decoherence. The scale factor $a$ of a 
Friedmann universe then becomes the relevant variable and attains the classical 
character through continuous measurement. In the literature a question arises 
that if the states $\exp{(iS)}$ and $\exp{(-iS)}$ describing an expanding and 
collapsing universe decohere, can one recover an approximate time-dependent 
Schroedinger equation from the timeless Wheeler-DeWitt equation and what are 
the boundary conditions at small scales [17] that lead to quantum effects in 
the vicinity of the turning point. In this work we answer all three questions. 
We show that (i) an approximate time dependent Schroedinger equations follows 
irrespective of the Wheeler-DeWitt equation, (ii) the adiabatic Gaussian 
wavefunction for the SWD equation is consistent with the Hartle-Hawking 
criterion plus the normalization prescription or the wormhole dominance proposal. 
This solves the problem of small scale boundary condition. 
(iii). Through the wormhole dominance proposal it has been explicitly shown that 
the wormholes lead to quantum effects in the vicinity of the turning point (in 
our case $a\simeq 0$) and classical turning point (in our case 
$a\simeq H^{-1}$) behaves as an starting point for the arrow of time and is 
manifested through the matter field as if the geometry itself looks into the 
evolution through the matter field. (iv). In the standard derivation of SWD 
equation, an expansion of the Wheeler-DeWitt wavefunction with respect to the 
Planck mass leads to difficulties as discussed in [18]. Our derivation is 
basically an expansion with respect to $\hbar$ and is justified through 
equations (30) and (31). (iv). The decoherence mechanism in our approach is the 
same as in ref. [3] and we do not repeat it here. (v). We also applied our 
approach to Starobinsky type $R^2$-cosmology  and find that the Gaussian 
ansatz correctly reproduces the wavefunction corresponding to the wormhole 
dominance proposal.
\smallskip
\begin{center}
{\bf{References}}
\end{center}
{\obeylines\tt\obeyspaces{ 
1. J.J.Halliwell and S.W.Hawking, {\it{Phys.Rev.}}{\bf{D31}}, 1777(1985)
2. C.Kiefer, {\it{Class. Quantum. Grav.}}{\bf{4}}, 1369(1987)
3. C.Kiefer, {\it{Phys. Rev.}}{\bf{D46}}, 1658(1992)
4. C.Kiefer, {\it{Phys. Rev.}}{\bf{D45}}, 2044(1992)
5. S.W.Hawking, {\it{Nucl. Phys.}}{\bf{B239}}, 257(1984)
    \quad S.W.Hawking and D.N.Page, {\it{Nucl. Phys.}}{\bf{B264}}, 184(1986)
6. A.Vilenkin, {\it{Phys. Rev.}}{\bf{D37}}, 888(1987)
7. S.Biswas, B.Modak and D.Biswas, {\it{Phys. Rev.}}{\bf{D55}},
    \quad 4673(1996)
8. D.N.Page,{\it{In Quantum concepts in space and time}}, ed. by R.Penrose 
    \quad and C.J.Isham (Clarendon, Oxford, 1986), p274.
9. A.A.Starobinsky, {\it{Phys. Lett.}}{\bf{B91}}, 99(1980)
10. G.C.Ghirardi, {\it{Phys. Rev.}}{\bf{D34}}, 470(1986)
11. D.Giulini, E.Joos, C.Kiefer, J.Kupsch, I.O.Stamatescu And 
    \quad H.D.Zeh, {\it{Decoherence and the Appearance of a Classical 
    \quad World in Quantum Theory}} (Springer, Berlin, 1996) 
12. C.Kiefer, D.Polarski and A.A.Starobinsky, {\it{gr-qc}}/9802003(1998) 
13. C.J.Isham, {\it{in: Integrable Systems, Quantum Groups, and Quantum
    \qquad Field Theories}}, by L.A.Ibart and M.A.Rodrigues, (Kluwer,
    \qquad Dordrecht, 1992), p 157-288, 
14. K.V.Kuchar, {\it{in : Proceeding of the fourth Canadian Conference 
    \quad on General Relativity and Relativity Astrophysics}}, 
    \quad ed. by Kunstatter G Vincent D Williams J 
    \quad (World Scientific, Singapore), p 211-314 (1992)
15. E. Madelung, Z. Phys. {\bf{40}}, 332(1926)
16. D. Bohm, Phys. Rev. {\bf{85}}, 166(1952)
17. H.D.Conradi and H.D.Zeh, {\it{Phys. Lett.}}{\bf{A154}}, 321(1991)
18. Betoni et. al, {\it{Class. Quant. Grav.}}{\bf{13}}, 2375(1996).}}
\smallskip
\section{\bf{Acknowledgment}}
A. Shaw acknowledges the financial support from ICSC World Laboratory, LAUSSANE
during the course of the work.
\end{document}